# Submilisecond acoustic pulses: effective pitch and Weber-Fechner law in discrimination of duration times

Marcin Majka[1], Paweł Sobieszczyk[1], Robert Gębarowski[2] and Piotr Zieliński[1,2]

[1]*The H. Niewodniczański Institute of Nuclear Physics, PAN, ul. Radzikowskiego 152, 31-342 Kraków, Poland*
[2]*Cracow University of Technology, Institute of Physics, ul. Podchorążych 1, 30-084 Kraków, Poland*
*e-mail address: Marcin.Majka@ifj.edu.pl*

The enclosed tests demonstrate that an effective pitch can be attributed to acoustic signals shorter then tenths of milliseconds. A power-law dependence of this pitch on the signal's duration time is found for subjects tested with Gaussian pulses. The discrimination threshold for the pulse duration time reported on the basis of the effective pitch increases proportionally to the duration time itself, i.e. it follows the Weber-Fechner law. A model based on the "Helmholtz's harp" idea, i.e. a series of damped resonators tuned in the audible range of frequencies reveals the mechanism of producing a maximum in the filtered spectrum of the pulse and corroborates the power law in the dependence of the position of the maximum on the duration time of the pulse. The model indicates a possibility of a manmade device designed to determine durations so short that they are inaccessible by direct measurements.

The discrimination threshold or difference limen, also known as the just noticeable difference, in the perception of physical stimuli by humans is in many cases proportional to the intensity of the stimulus itself. This observation made by E. H. Weber [1] is conveniently summarized as proportionality of the intensity of the perceived sensation to a logarithm of the intensity of the stimulus. The statement is called Weber-Fechner law [2,3]. Although often limited to a certain range of magnitude of the stimulus, and in some cases replaced by different functional stimulus-sensation relations [2], the Weber-Fechner law constitutes a natural base for logarithmic scales used in many applications, e.g. in determination of the distance of a star by its stellar magnitude [4] or in the decibel scale of the sound intensity level.

The physics of short and, consequently, broad-band signals is now a growing research subject. Femtosecond electromagnetic pulses produced by free electron laser (FEL) sources are used in studies of evolution of electronic states in materials [5-10]. Experiments on the acuity in perception of the pitch (i.e. a logarithm of frequency) and the time localization of short acoustic pulses have revealed extraordinary capabilities of the human hearing apparently beyond limitations imposed the Heisenberg's uncertainty relation [11-13]. The pulses studied in [13-19] were, however, long enough to speak legitimately of an enveloped monochromatic wave. Electromagnetic pulses shorter than one oscillation cycle are now available [20] and some aspects of their interaction with matter have been already studied. [21]. In a comment [22] to ref. [11] the present authors found, that an approximate, effective pitch is perceived for pulses with time envelopes far too short to allow the fundamental frequency to be defined. Generally, a decrease in duration of the pulse resulted in an increase in the effective pitch [22].

In what follows we give the effect a quantitative description for acoustic Gaussian submilisecond pulses, although the results may be useful in spectral analysis of wave packets of any nature. The time extent of the pulse is defined by the width parameter $\sigma$ (standard deviation) of the Gaussian profile of acoustic pressure. A simple test has been constructed to examine the just perceptible difference in the signal duration $\sigma$ by humans. The subject is presented consecutive pairs of sounds. The width parameters are: $\sigma$ in the first, and $\sigma + \Delta\sigma$ in the second sound. The value $\Delta\sigma$ in the consecutive pairs is progressively incremented starting from zero. The subject is asked to indicate the difference limen value $\Delta\sigma_l$ i.e. the one that produces the first perceived difference between the sounds of a pair. The same test has been also performed in the reverse order i.e. starting from large width differences $\Delta\sigma$. Then, the subjects were asked to notice the value $\Delta\sigma_L$ at which the sounds start to be perceived as identical.

The results of four subjects (authors) are reported in Fig. 1. MM has an absolute hearing whereas PS, PZ, RG have relative hearing. MM, PS and PZ have a practice in choir singing, MM additionally in violin playing and PZ in piano playing. RG is a music lover with, however, no systematic ear training. {Tests are described in more detail in the supplementary material [23]}

The responses of the ear-trained subjects are clearly less scattered and their reported difference limens lie systematically lower than those noticed by the untrained subject. It is, however, interesting that MM and RG are able to distinguish two different qualities of the sound, i.e. its pitch and timbre, (comp. sharpness [17]) separately. Although the responses of both subjects are rather different in absolute values the behavior of the quality termed timbre as a functions of the width $\sigma$ is quite analogous. The difference limen in the perceived pitch increases with increasing $\sigma$, whereas the limen in timbre is practically independent of $\sigma$. On the other hand, the subjects with relative hearing and ear training (PS and PZ) seem to perceive the difference in $\sigma$ by a composed or united quality encompassing both the pitch and the timbre. This may result from a practice or even an imprinted compulsion of attributing a musical interval to every pair of the sounds heard.

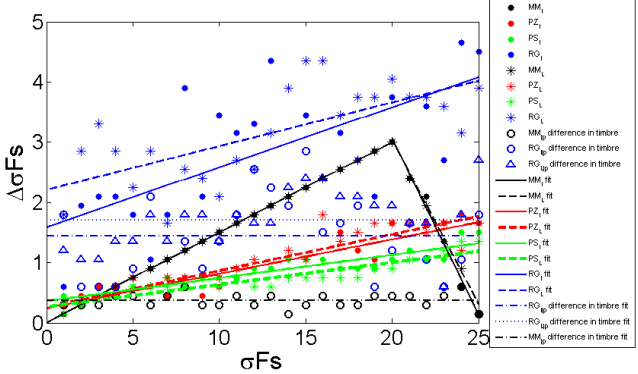

Fig. 1. Results of test on discrimination of width parameter of Gaussuan pulses by 4 subjects. Index l corresponds to increasing and L to decreasing change in the duration of second sound.

The just perceptible difference in the effective pitch as reported by MM follows the Weber-Fechner law up to about $\sigma = 20/F_s \approx 0.45 \cdot ms$ (where the sampling rate $F_s = 44100 \text{ s}^{-1}$), i.e.

$$\Delta \sigma_{l,P} = A_0 + A_1 \sigma, \quad (1)$$

with $A_0 = 0$, where $\Delta \sigma_{l,P}$ denotes the just perceptible difference $\Delta \sigma_l$ in the pulse width as perceived by the corresponding difference in the effective pitch. The threshold on the discrimination of the pulse width $\Delta \sigma_{l,S}$ on the basis of the sole sound's timbre turns out practically independent of the initial width. This indicates that humans possessing the capability of distinguishing both qualities perceive, by the differences in timbre, an absolute and not relative value of the width change, i.e.

$$\Delta \sigma_{l,S} = C. \quad (2)$$

In the light of the above results the responses of the listeners who perceive a joint, united effect "U" of the pitch and timbre together can be understood as a certain average of both "pure" qualities. Within such a hypothesis and under the assumption that the results of MM are most reliable, i.e. with the use of the coefficients $A_1$ and $C$ for MM the responses of PS and PZ can be fitted to straight lines

$$\Delta \sigma_{l,U} = aA_1\sigma + bC. \quad (3)$$

with some coefficients $a$ and $b$. All the fitted straight lines are shown in Fig 1. The coefficients of eqs (1), (2) and (3) for all the subjects are gathered in Table 1.

Table 1.

| Subject | $A_0$ [s/44100] | $A_1$ [s/44100] | $C$ [s/44100] | $a$ | $b$ | $\alpha$ |
|---|---|---|---|---|---|---|
| MM$_l$ | 0.00±0.01 | 0.15±0.01 | 0.375±0.115 | 1.00±0.07 | 0.00±0.01 | 1.30±0.05 |
| MM$_L$ | 0.00±0.01 | 0.15±0.01 | - | 1.00±0.07 | 0.00±0.01 | - |
| RG$_l$ | 1.59±0.34 | 0.100±0.023 | 1.44±0.65 | - | - | - |
| RG$_L$ | 2.21±0.24 | 0.072±0.016 | 1.71±0.48 | - | - | - |
| PS$_l$ | - | - | - | 0.255±0.022 | 0.96±0.13 | 0.271±0.031 |
| PS$_L$ | - | - | - | 0.245±0.021 | 0.72±0.13 | - |
| PZ$_l$ | - | - | - | 0.38±0.03 | 0.64±0.19 | 0.260±0.023 |
| PZ$_L$ | - | - | - | 0.40±0.04 | 0.69±0.24 | - |

Table 1. Parameters of eqs. (1), (2), (3) and (4) as fitted for the subjects examined.

It is known that the discrimination of frequencies of long periodic signals by humans follows the Weber-Fechner law $\Delta \omega_l / \omega = const = G$, which is best represented by a logarithmic frequency dependence of the musical pitch [24]. If the effective pitch of a pulse $\omega_{EP}(\sigma)$ (Effective Envelope Pitch) also follows such a rule the Weber-Fechner law for the discrimination of width $\sigma$ implies the power law relation

$$\omega_{EP}(\sigma) = \frac{D}{\sigma^\alpha}. \quad (4)$$

To verify the hypothesis of eq. (4) the subjects MM, PS and PZ compared the pitch attributed to Gaussian pulses of standard deviation $\sigma$ with longer reference sounds of easily discernible pitches. The effective pitch was assessed by attributing a musical interval between the reference sound and the pulse (see Supplemental material [23] for details) The subject without ear training RG was not examined as he had problems in defining intervals. Fig. 3 shows a Log-Log plot of the Envelope's Effective Pitch $\omega_{EP}(\sigma)$ for 3 subjects. The results corroborate the power law behavior of eq. (4) in the range $1/F_s < \sigma < 25/F_s$ (see table 1 for fitted values of the exponent). The exponents $\alpha$ for low $\sigma$ differ significantly in the subject capable of distinguishing the pitch and the timbre separately. The likeness of the exponent $\alpha$ and the existence of a region $25/F_s < \sigma < 50/F_s$ of a steeper descent in the subjects with relative hearing are remarkable.

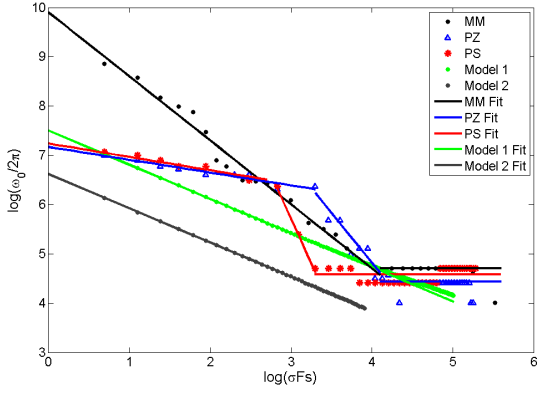

Fig. 3. Log-Log plot of Envelope's Effective Pitch $\omega_{EP}(\sigma)$ attributed to Gaussian pulses of width $\sigma$ for 3 subjects. Shown are also positions of maxima of amplitude obtained with model based on series of damped oscillators, with $\tau_r = 0.002$ s (Model 1) and with empirical $\tau_r$ [25] (Model 2).

To reveal possible mechanisms of the above experimental findings we have constructed a model of the frequency perception based on the Helmholtz's idea of a "harp". It consists of a series of resonators each tuned to its own eigenfrequency $\omega_0$ and damped with a damping time $\tau_r$ [26,27]. The impulse response function of such a resonator is known $K(t;\omega_0,\tau_r) = \theta(t) \frac{\exp(-t/\tau_r)}{\omega_r} \sin(\omega_r t)$, where $\omega_r = \sqrt{\omega_0^2 - 1/\tau_r^2}$ .

A drawback of such a generic model is that response systematically grows with decreasing frequency. To avoid this behavior we assume that the low Fourier components reaching the corresponding resonators are additionally weakened as it follows from more distant places of the resonators with respect to the oval window of the cochlea [28]. This effect is accounted for in our model by a simple multiplication of the response function by the eigenfrequency $\omega_0$.

$$\tilde{K}(t;\omega_0,\tau_r) = \omega_0 \frac{\theta(t)\exp(-t/\tau_r)\sin\left(\sqrt{\omega_0^2 - 1/\tau_r^2}\, t\right)}{\sqrt{\omega_0^2 - 1/\tau_r^2}}. \quad (8)$$

The response to a given signal

$$s(t;\sigma) = \exp(-t/2\sigma^2). \quad (9)$$

at the resonator $\omega_0$ then is

$$u(\omega_0,t) = \int_{-\infty}^{t} \tilde{K}(t-t';\omega_0,\tau_r)s(t')dt'. \quad (10)$$

We attribute the effective perceived pitch of the Gaussian signal of width $\sigma$ to the maximum of the function $u(\omega_0,t)$ with respect to time and frequency. A precise detection of the position of such a maximum is beyond the scope of this paper. In the cochlea of living mammals the corresponding mechanism involves nonlinear cochlear tuning due to complex motions of the outer hair cells [29-31]. Alternative and/or synergic mechanical effects have also been evoked [32]. The plot of the resonant frequency $\omega_r$ corresponding to the maximum response (eq. (10)) as a function of the width $\sigma$ for a constant damping time $\tau_r = 0.002$ s is given in Fig. 3. In spite of its simplicity the model reproduces the power law of Eq. (4) in the region studied with $\alpha = 0.6939 \pm 0.0005$. The exponent is robust to variations of the damping time $\tau_r$ in a large range. Too short damping times narrow the range of linearity in the log-log plot. Putting the damping time to an empirical estimate of the threshold on the temporal gaps detection in narrow-band noise [25] $\tau_r = 25.67 - 6.33\log_{10}(\omega_0/2000\pi)$ [ms] one obtains practically the same exponent $\alpha = 0.6951 \pm 0.0008$, with, however, the effective pitch much below that reported by subjects. The comparison of this result with our experimental data in Fig. 3 indicates that additional filtering takes place due to anatomical differences and training of the subject and/or to the quality of the sound sources and of the rooms where the tests were carried out. It is, however, clear that the main mechanism responsible for the existence of the effective envelope pitch relies on a band pass filter that cuts a weighted part of the spectrum of the pulse in such a way that that the filtered spectrum shows a maximum. To verify this hypothesis we are now going to extend our studies beyond the Gaussian pulses.

It still remains an open question which haracteristics of the so filtered spectrum is responsible for the sensation of timbre surprisingly reported by ear non-trained subject. It is plausible to attribute it to an integral quantity of the spectrum. The first candidate is the modulus of the difference of the envelopes for neighboring widths integrated over frequency. One can easily realize that the non filtered envelope of the pulse given in eq. (9), i.e. normalized to the peak height in the time domain is normalized to the area under the pulse in the frequency domain. Then the integrated difference modulus in the limit of small width differences is inversely proportional to the width itself $b^{(1)} = \left[\int_0^\infty \left|\frac{\partial s(\omega,\sigma)}{\partial \sigma}\right| d\omega\right] d\sigma = \sqrt{\frac{2}{\pi e}} \frac{1}{\sigma} d\sigma$.

If the sensation is produced by the quantity $b^{(1)}$ then the Weber-Fechner law is satisfied. Thus, it would be rather an alternative candidate for the sensation of the effective pitch. On the other hand, the integrated modulus of differences of the power spectra of the neighboring pulses is independent of $\sigma$

$b^{(2)} = \left[\int_0^\infty \left|\frac{\partial |s(\omega,\sigma)|^2}{\partial \sigma}\right| d\omega\right] d\sigma = \frac{d\sigma}{2\sqrt{\pi}} \left(\text{erf}(1) + 2e^{-1}/\sqrt{\pi} - 1/2\right)$.

This indicates that the power spectrum might be responsible for the sensation of the difference in the pulse's timbre.

We have checked the $\sigma$ dependence of analogous quantities in the spectra obtained as maxima of the response of eq. (10). The results are plotted in Figs 4 and 5.

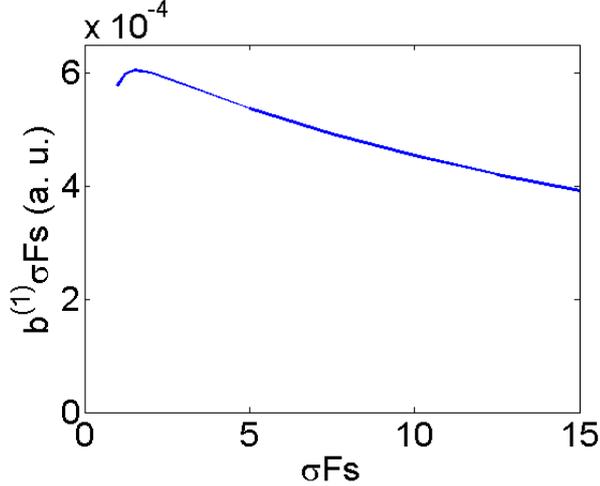

Fig. 4 Quantity $b^{(1)}\sigma$ for the model with $\tau_r = 0.002$ s.

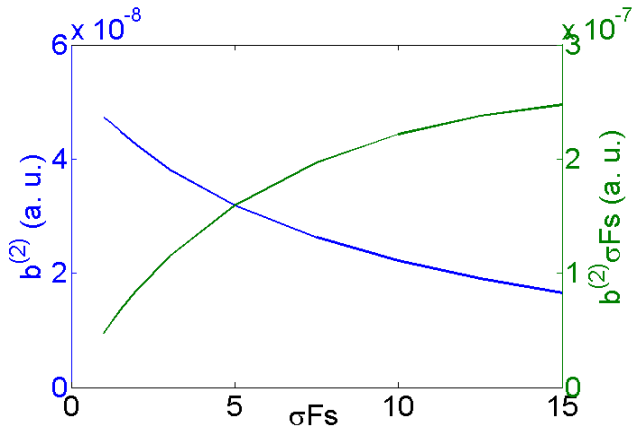

Fig. 5 Quantity $b^{(2)}$ (blue) and $b^{(2)}\sigma$ (red) for the model with $\tau_r = 0.002$ s.

The results are not unambiguous. While $b^{(2)}$ is not entirely constant its decrease is slower than that implied by the Weber-Fechner law. A cross over behavior visible at small widths in Fig. 4 originates from the range of integration limited to the audible frequency region. A stronger filtering out of high frequencies will result in an enlargement of this region that can mimic a constant behavior.

In conclusion, we have shown that ultrashort acoustic pulses produce an effective pitch related to the pulse width by a power law. An appropriate filtering, here with the use of a harmonic model based on the Helmholtz "harp idea", has turned out to provide a spectrum with maximum that reproduces qualitatively the power law behavior. The Weber-Fechner law for the discrimination of widths of Gaussian pulses gives a powerful tool of the assessment of the pulse duration far beyond the possibilities of a direct insight or measurement. This is an analogy to diffraction methods that allow one to determine interatomic distances from the information in reciprocal space. The Weber-Fechner discriminability based on the effective pitch works most efficiently for the shortest pulses. It may indicate a method of measuring parameters of short pulses in other areas of physics.

**Acknowledgment**
Two of us (MM and PS) acknowledge support by Krakowskie Konsorcjum "Materia-Energia-Przyszłość" im. Mariana Smoluchowskiego as a part of KNOW 2012-2017, KKS123/8-09985.

**References**

[1] D. M. Green, *J. Acoust. Soc. Am.* **32**, 121-131 (1960).
[2] B. C. J. Moore, *An Introduction to the Psychology of Hearing*, Brill, Leiden, (2013) p. 144
[3] P. A. van der Helm, *Attention, Perception & Psychophysics* **72** (7), 1854-1864 (2010).
[4] J. Dufay, *Introduction to Astrophysics: The Stars*, Dover Publications, New York, (2012).
[5] T. Rohwer et al., *Nature* **471**, 490–493 (2011).
[6] S. Hellmann et al., *Phys. Rev. Lett.* **105**, 187401 (2010).
[7] J. Andruszkow et al., *Phys. Rev. Lett.* **85**, 18 (2000).
[8] J. T. Costello, *Journal of Physics: Conference Series* **88**, 012057 (2007).
[9] S Hellmann et al., *New Journal of Physics* **14**, 013062 (2012).
[10] A Pietzsch et al., *New Journal of Physics* **10**, 033004 (2008).
[11] J. N. Oppenheim and M. O. Magnasco, *Phys. Rev. Lett.* **110**, 044301 (2013).
[12] D. A. Ronken, *J. Acoust. Soc. Am.* **49**, 1232 (1971).
[13] B. C. J. Moore, *J. Acoust. Soc. Am.* **54**, 610 (1973).
[14] M. O. Magnasco, *Phys. Rev. Lett.* **90**, 058101 (2003).
[15] R. F. Lyon and M. A. Carver, *Hydrodynamics Demystified*, California (1988). Unpublished
[16] H. Duifhuis, *J. Acoust. Soc. Am.* **48**, 888 (1970).

[17] R. A. Fearn, *Music and pitch perception of cochlear implant recipients*, New South Wales (2001).
[18] T. D. Rossing and A. J. M. Houtsma, *J. Acoust. Soc. Am.* **79**, 1926 (1986).
[19] P. Mohlin, *J. Acoust. Soc. Am.* **129**, 3827 (2011).
[20] H.-C. Wu and J. Meyer-ter-Vehn, *Nature Photonics* **6**, 304–307 (2012).
[21] R. Gębarowski, *J. Phys. B: At. Mol. Opt. Phys.* **30**, 2143–2154 (1997) .
[22] comment to ref [11] submitted to *Phys. Rev. Lett*
[23] See Supplemental Material at
[24] A. Houtsma, in *Hearing, Handbook of Perception and Cognition*, B.C.J. Moore Ed.  Academic, San Diego, (1995) p. 267
[25] P. J. Fitzgibbons, *J. Acoust. Soc. Am.* **74**, 1 (1983).
[26] H. Helmholtz, *On the sensations of tone*. Dover Publications, New York, (1954).
[27] E. L. LePage, in *Lecture Notes in Biomathematics - The Mechanics and Biophysics of Hearing*, P. Dallos et al. Eds. Springer, Berlin, (1990) p. 278-287
[28] A. M. Gilroy *et al.*, *Atlas of anatomy*, Thieme, (2008) p. 536
[29] J. A. N. Fisher *et al., Neuron* **76**, 5 (2012).
[30] P. Dallos *et al., Neuron* **58**, 3 (2008).
[31] A. Sęk *et al.*, *Int J Audiol*. **44**, 408-420 (2005).
[32] V. Martinez-Eguíluz *et al.*, *Phys. Rev. Lett.* **84**, 5232 (2000).

# Supplemental Material for Paper
## "Submilisecond acoustic pulses: effective pitch and Weber-Fechner law in discrimination of duration times"

**Instructions for the use of acoustic tests**

The sound files described here are available on the web sites cited in [1,2,3].

**I Test for discrimination of duration times of Gaussian pulses**

The reader can check the discrimination limens for the width parameter $\sigma$ of Gaussian signals $\exp(-t^2/2\sigma^2)$ in the range $1/F_s < \sigma < 25/F_s$, i.e. $0.023 \text{ ms} < \sigma < 0.567 \text{ ms}$, where $F_s = 44100 \text{ s}^{-1}$ is the standard sampling rate in the *.wav format of sounds. The sound files are to be found at the site, ref. [1]. There are 25 files, each corresponding to the initial width parameter $\sigma = NN/F_s$ where the integer number NN is indicated in the file name dlNN.mp4. Each file contains a series of pairs of sounds.  The first sound has the width parameter $\sigma = NN/F_s$ and the second $(XX + n \cdot \Delta\sigma)/F_s$, where $\Delta\sigma = 0.15/F_s$. The number *n* is seen in the screen. The listener is asked to indicate the pair number $n_l$ for which the sounds appear to be different. The corresponding difference $n_l \cdot \Delta\sigma/F_s = \Delta\sigma_l$ is the discrimination limen sought. If the discrimination limen is proportional to the initial width $NN/F_s$ $\sigma = NN/F_s$ then the Weber-Fechner law is satisfied. Unexpectedly, the responses of two authors (MM and RG) distinguished two different qualities of the sounds: pitch and timbre [4,5]. Whereas the difference limen $\Delta\sigma_{l,P}$ observed on the basis of the pitch followed the Weber-Fechner law, the discriminability of the width as perceived through the timbre $\Delta\sigma_{l,S}$ were statistically independent on the initial width parameter. Other authors (PS and PZ) reported only one quality that allowed them to distinguish the sounds, and the Weber-Fechner law was fairly well fulfilled as it is shown in Fig. 1 in the main text. The test [2] is analogous, but the difference in the width parameter $n \cdot \Delta\sigma/F_s$ decreases starting from $n = 30$ to $n = 0$. The listener may then find the discrimination limens $\Delta\sigma_{L,P}$, $\Delta\sigma_{L,S}$ or just $\Delta\sigma_L$. All the pulses are normalized to equal amplitudes of the Gaussian peak. Some subjects (not reported here) notice a difference in loudness.

**II Test for Effective Envelope Pitch of a Gaussian pulse**

This test is designed mainly for the listeners with relative pitch hearing. The file in the eep.mp4 is available at the site given in ref [3]. The file consists of pairs of sounds. The first sound is a cosine wave of frequency equal 660 Hz enveloped by a Gaussian of a width parameter equal to $1000/F_s$. The pitch of this reference sound is easy to determine by an average listener. The second sound is a pure Gaussian envelope $\exp(-t^2/2\sigma^2)$. The maximum amplitudes of signals are equal. The width $\sigma$ of the Gaussian decreases in the progress of time of the experiment in the following manner: *i*) starting from $1000/F_s$ to $200/F_s$ by $50/F_s$ and *ii*) from $199/F_s$ to $1/F_s$ by $1/F_s$. The actual width of the pulse is seem in the screen. The listener, who is able to note the relative pitch and has some knowledge of musical intervals can attribute an interval to each pair heard. Then the frequency $\omega_{EP}(\sigma)$, i.e. the one defining the Effective Envelope Pitch should be calculated using the known frequency ratio of each interval. Listeners with absolute hearing may find the reference sound redundant. The listeners without any knowledge of musical intervals may give their responses in terms of beginnings of known tunes. Multiple listening to the test may give more precise results. Sometimes subjects hesitate on the attribution of a perceived interval due to an approximate character of the Effective Envelope Pitch. The responses may also be different for subjects who are used to different intervals than those known in the European music. Some subjects (not reported here) can notice differences in loudness between the reference sound and the Gaussian pulse as a disturbing factor in recognizing the intervals.

**References:**

[1]     https://drive.google.com/folderview?id=0BwQPqgssghTtTWxZcG8wckhQcGc&usp=sharing
[2]     https://drive.google.com/folderview?id=0BwQPqgssghTtdlVsR00zd1lxNTg&usp=sharing
[3]     https://drive.google.com/file/d/0BwQPqgssghTtT041dWpwYUpZRjg/view?usp=sharing
[4]     B. J. C. Moore**,** An Introduction to the Psychology of Hearing, Brill, Leiden, (2013),  defines  timbre as an attribute of sound (p. 89), and as spectral shape  (p. 109) the chapter on pitch pitch (p. 3)
[5]     C. J. Plack, A. J. Oxenham, R. R. Fay and A. N. Popper, *Pitch: Neural Coding and Perception.* New York: Springer (2005).